\documentclass[10pt,a4paper]{article}

\usepackage[a4paper,top=3.3cm,bottom=4.2cm,left=2.6cm,right=2.6cm]{geometry}
\usepackage[T1]{fontenc}
\usepackage[utf8]{inputenc}
\usepackage{newtxtext,newtxmath}
\usepackage{graphicx}
\graphicspath{{media/media/}}
\usepackage{booktabs}
\usepackage{array}
\usepackage{multirow}
\usepackage{caption}
\usepackage{setspace}
\usepackage{titlesec}
\usepackage{enumitem}
\usepackage{amsmath}
\usepackage{hyperref}
\usepackage{float}

\titleformat{\section}[hang]
  {\normalfont\bfseries\fontsize{13}{15}\selectfont}
  {\makebox[0.7cm][l]{\thesection}}
  {0pt}
  {}
\titlespacing*{\section}{0pt}{24pt}{12pt}

\titleformat{\subsection}[hang]
  {\normalfont\bfseries\fontsize{12}{13}\selectfont}
  {\makebox[1.0cm][l]{\thesubsection}}
  {0pt}
  {}
\titlespacing*{\subsection}{0pt}{11pt}{11pt}

\titleformat{\subsubsection}[hang]
  {\normalfont\bfseries\fontsize{11}{12}\selectfont}
  {\makebox[1.0cm][l]{\thesubsubsection}}
  {0pt}
  {}
\titlespacing*{\subsubsection}{0pt}{10pt}{10pt}

\newcommand{\keywords}[1]{%
  \vspace*{24pt}
  {\fontsize{9}{11}\selectfont\noindent
  \makebox[2cm][l]{Keywords:}%
  \parbox[t]{\dimexpr\linewidth-2cm\relax}{#1.}\par}
}

\newcommand{\abstracttext}[1]{%
  \vspace*{12pt}
  {\fontsize{9}{11}\selectfont\noindent
  \makebox[2cm][l]{Abstract:}%
  \parbox[t]{\dimexpr\linewidth-2cm\relax}{#1}\par}
  \vspace*{30pt}
}

\newcommand{\firstpar}[1]{\noindent #1}

\begin{document}

\begin{center}
{\fontsize{15}{17}\selectfont\bfseries LLM-Based Synthetic Ground Truth Generation for Audio-Based Emotion Classification via In-Context Learning\par}
\vspace*{24pt}
{\fontsize{11}{13}\selectfont Qing Huang\textsuperscript{1,2}, Pooja Pol\textsuperscript{1,2} and Jianing Zhang\textsuperscript{1}\par}
\vspace*{12pt}
{\fontsize{9}{11}\selectfont\itshape
\textsuperscript{1}School of Business, Technical University of Applied Sciences Augsburg, Friedberger Str. 4, 86161 Augsburg, Germany\par
\textsuperscript{2}Data Science und Autonome Systeme Technologietransferzentrum (TTZ), An der Schmiede 19, 86899 Landsberg am Lech, Germany\par
qing.huang@tha.de, pooja.pol@tha.de, jianing.zhang@tha.de\par}
\end{center}

\keywords{Large Language Models (LLMs), In-Context Learning (ICL), Synthetic Ground Truth, Affective Computing, Data-Driven Decision Support, Virtual Reality (VR)}

\abstracttext{Understanding human states and interaction dynamics is a core goal of human--computer interaction (HCI). As interaction paradigms become more immersive, virtual reality (VR) has emerged as a powerful platform for studying collaborative work. In such settings, evaluating team collaboration states, including team performance and team resilience, requires continuous and reliable inference of latent team-level cognitive and affective states from multi-modal sensor data, such as speech signals. However, generating ground truth labels for these latent states remains challenging due to sensor-induced noise, contextual variability, and sparse expert annotations. Traditional self-reporting approaches provide only static and delayed measurements and are therefore insufficient for capturing dynamic team processes reflected in continuous speech data. In this work, we propose a large language model (LLM)-driven, agentic inference workflow for automated emotion-related synthetic ground truth generation from streaming speech data in multi-user VR environments. Leveraging the generalization capabilities of LLMs, we use In-Context Learning (ICL) with few-shot demonstrations of paired audio-based samples and their corresponding transcriptions. ICL tends to achieve task adaptation comparable to model fine-tuning while circumventing the computational overhead of parameter updates. To construct informative and robust in-context prompts, we adopt a retrieval-based selection strategy that dynamically identifies relevant audio demonstrations based on similarity in the acoustic feature space. Specifically, retrieval is guided by prosodic descriptors derived from speech signals, whereas transcript-based semantic information is incorporated directly within the prompt and interpreted by the LLM through its reasoning capabilities. This modality-aware strategy promotes consistent affective labeling and improved generalization across previously unseen speech segments. Evaluated on multi-player VR audio recordings, our methodology demonstrates potential as a scalable, data-efficient component for data-driven team-based decision support. By integrating acoustic similarity-based retrieval with LLM-based semantic reasoning, this work contributes to emerging interdisciplinary methodologies at the intersection of scientific machine learning, multi-modal systems, and AI-driven decision-making.}

\section{INTRODUCTION}
\firstpar{Virtual reality (VR) provides an immersive platform for analyzing human behavior through multimodal sensing, including speech, motion, and physiological signals~\cite{ref7}. In collaborative VR tasks, speech plays a central role, reflecting coordination, intent, and affective expression. However, establishing reliable ground truth annotations for affective states remains challenging. Self-report measures are subjective and temporally misaligned~\cite{ref6}, while sensor data are noisy and large-scale, making manual annotation costly. These constraints motivate automated and data-efficient annotation strategies.}

We focus on speech sentiment annotation in collaborative VR environments. Recent advances in large language models (LLMs) enable affective reasoning via in-context learning (ICL) without task-specific parameter updates, often approaching fine-tuned performance~\cite{ref2,ref8}. This makes LLMs particularly suitable for scalable annotation scenarios where labeled data are limited and repeated manual labeling across sessions is costly.

In collaborative VR settings, sessions involve changing player constellations while task contexts and discussion structures remain comparable. Although each session includes only 5--7 participants, affective expressions are influenced not only by individual speaker characteristics but also by shared interactional dynamics emerging from common tasks and conversational context. For consistent annotation across sessions, it is therefore beneficial to align segments at the level of comparable affective interaction patterns that arise under similar task conditions. To support such cross-session consistency, our approach retrieves acoustically comparable segments using generic prosodic descriptors (e.g., pitch, loudness, intensity, and speaking rate). These descriptors capture expressive characteristics that remain relatively stable across different speaker constellations and thereby provide a session-independent basis for demonstration selection.

Building on this motivation, we propose an LLM-based framework for synthetic speech sentiment annotation that integrates speech-oriented LLMs with a modality-aware retrieval-guided ICL strategy grounded in acoustic similarity. While ICL and retrieval-based prompting are established techniques, the novelty of our approach lies in positioning acoustic similarity as the primary mechanism for selecting affectively comparable interaction segments across sessions, thereby enabling context-level alignment rather than session-specific adaptation. Retrieval is conducted in the acoustic feature space to identify segments that share comparable expressive patterns, while transcript-based content is incorporated within the prompt and interpreted by the LLM through its semantic reasoning capabilities. Rather than introducing a new similarity metric per se, we reconceptualize retrieval-guided ICL as a structured mechanism for cross-session affective alignment. This design mitigates over-adaptation to session-specific characteristics and instead supports data-efficient domain adaptation based on shared interactional dynamics. Because retrieval relies on generic acoustic descriptors and semantic interpretation is handled by the LLM, the framework generalizes beyond a particular team composition and can be transferred to other collaborative or conversational settings without model fine-tuning.

\section{DATA ACQUISITION}
\firstpar{This study is based on data collected from a multi-player VR game in which participants collaborate to achieve a shared objective. The VR scenario is designed to elicit natural team interaction and communication under controlled yet immersive conditions, enabling analysis of individual- and team-level behavioral dynamics.}

Data were collected across two VR sessions. The first session served as a familiarization phase, during which participants explored the virtual environment and learned the task mechanics, while the second session focused on task execution, with participants collaboratively completing the assigned task. This session-based design enables the capture of multi-modal sensor data under different interaction conditions and supports analysis of condition-dependent variations in cognitive and affective states.

Each session involved approximately five to seven participants equipped with VR head-mounted displays (Meta Quest Pro), allowing continuous recording of multi-modal sensor data. Speech audio was recorded in a single-channel (mono) format via the headset-mounted microphone array, which provides a downmixed audio signal, and encoded on device at a sampling rate of 48~kHz. Each participant's audio recording has an average duration of approximately 25 minutes.

In this work, we focus on speech signals recorded during team interactions. The collected audio data capture spontaneous verbal communication and are predominantly in German, reflecting realistic collaborative settings. While most recordings were conducted in German, the proposed analysis and annotation framework is not inherently language-specific and can, in principle, be adapted to other languages.

For downstream analysis, the audio streams were processed using an automatic speech recognition (ASR) model based on Whisper~\cite{ref9} to generate textual transcriptions with preserved temporal alignment between audio segments and text. This alignment reduces potential temporal mismatches during affective annotation. The resulting paired audio--text data form the basis for synthetic sentiment annotation and further analysis.

For lexicon-based methods relying on the NRC-VAD resource, which is originally constructed for English, German transcripts were automatically translated into English prior to lexicon lookup. The translation step enables compatibility with the NRC-based affective scoring pipeline while preserving segment-level temporal alignment. In our analysis, cross-linguistic and cultural differences introduced by the translation step were observed to have only a limited impact on the overall annotation results. However, in our framework, lexicon-based annotations serve primarily as a lightweight reference signal rather than as the sole source of ground truth. Moreover, the retrieval-guided LLM-based annotation process operates directly on original transcripts and acoustic descriptors, thereby reducing dependence on language-specific lexical mappings. This design mitigates potential cultural bias introduced by cross-lingual lexicon transfer while maintaining methodological consistency.

\section{METHODOLOGY}
\firstpar{To enable automatic synthetic ground truth generation for speech sentiment annotation, we propose an LLM-based methodology built upon the publicly available Voxtral model~\cite{ref10}, a speech-oriented large language model capable of jointly processing audio and textual inputs. All experiments are conducted without task-specific fine-tuning. Inference is performed with fixed decoding parameters to reduce stochastic variability during generation. While different LLM backbones or decoding strategies may lead to quantitative differences in output, the proposed framework is designed in a modular manner and can, in principle, be instantiated with alternative speech-capable LLMs. A systematic comparison of different LLM backbones and their inference variability is beyond the scope of the present study and will be investigated in future work.}

We first established a vanilla ICL baseline using few-shot prompting with labeled audio examples. We then extend this baseline with a retrieval-guided ICL strategy that integrates acoustically derived representations for demonstration selection while incorporating transcript-based semantic information directly within the prompt for multimodal reasoning. Importantly, retrieval operates exclusively on precomputed acoustic descriptors and relies on similarity-based ranking without random sampling or approximate search. For a fixed ICL bank and identical input, the same set of in-context demonstrations is deterministically retrieved. As a result, retrieval stability is guaranteed by design, and any potential variability is confined to the generative stage of the LLM rather than the demonstration selection process.

\subsection{Vanilla ICL with Few-shot Prompting}
\firstpar{As a baseline approach for synthetic ground truth generation, we adopt the strategy demonstrated in Figure~\ref{fig:workflow}. In this setup, the model is prompted using a small number of labeled audio examples without any task-specific parameter updates. Each shot consists of a triplet of audio snippets, one per sentiment class (positive, negative, and neutral), paired with their corresponding sentiment labels provided in textual form. For each inference instance, a small number of such shots are randomly sampled and included in the prompt context to expose the model to representative sentiment variations. Following these few-shot demonstrations~\cite{ref8}, an unseen audio segment is appended to the prompt, and the model is instructed to infer its sentiment based on the contextual examples. This vanilla formulation enables the LLM to adapt to the task at inference time solely through contextual conditioning.}

\subsection{Acoustically Informed LLM-Based Synthetic Annotation via ICL}
\firstpar{Recent progress in self-supervised speech representation learning has demonstrated that pretrained acoustic models can effectively capture affective cues directly from raw audio signals. Transformer-based architectures such as wav2vec 2.0~\cite{ref1} and HuBERT~\cite{ref1} have been shown to predict continuous emotional dimensions, including valence, arousal, and dominance~\cite{ref4}, without relying on textual input. These properties make acoustic representations particularly valuable in spontaneous and conversational settings, where transcripts may be incomplete or noisy.}

In parallel, lexicon-based approaches provide an interpretable and lightweight means of extracting affective information from text. Resources such as the NRC word-emotion association~\cite{ref3} and NRC Valence, Arousal, and Dominance (NRC-VAD) lexicons~\cite{ref5} offer linguistically grounded affective cues that complement acoustically derived features and serve as useful reference annotations.

\begin{figure}[H]
    \centering
    \includegraphics[width=0.5\textwidth]{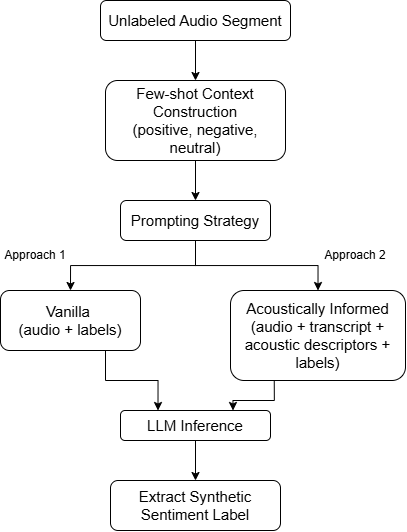}
    \caption{Flowchart illustrating the vanilla and acoustically informed ICL strategies.}
    \label{fig:workflow}
\end{figure}

Motivated by these observations, the proposed method integrates acoustic and semantic information into the ICL framework. Specifically, the acoustic information is represented using low- and mid-level speech descriptors, including pitch, loudness, intensity, and speaking rate measured in both words per second and syllables per second. As shown in Figure~\ref{fig:workflow}, a small set of few-shot demonstrations is enriched with both transcript-based content and acoustically derived descriptors, enabling LLM inference to be grounded in multimodal context. By combining the generalization capability of LLMs with acoustically informed ICL, the framework aims to improve the robustness and consistency of synthetic sentiment annotations, particularly in emotionally ambiguous conversational scenarios.

To systematically select informative demonstrations, we replace random sampling with similarity-based retrieval over an annotated segment pool $\mathcal{D}$.

For each segment $u_i$, we construct an acoustic descriptor vector
\begin{equation}
\mathbf{a}_i = E_{\mathrm{ac}}(x_i) \in \mathbb{R}^{d_a},
\end{equation}
where $E_{\mathrm{ac}}(\cdot)$ extracts normalized prosodic features (e.g., pitch statistics, loudness/intensity statistics, and speaking-rate measures) from the raw audio signal. Prior to retrieval, acoustic vectors are standardized to ensure comparability across feature dimensions.

Given a target segment $u^*$ with acoustic representation $a^*$, the acoustic distance is defined using Euclidean distance:
\begin{equation}
d_{\mathrm{ac}}(u^*,u_i) = \left\lVert \widetilde{a}^{*} - \widetilde{a}_{i} \right\rVert_2,
\end{equation}
where $\widetilde{a}$ denotes the standardized acoustic vector. Since smaller distances indicate higher similarity of acoustic features, the in-context demonstration set is obtained by selecting the top-$K$ segments with the smallest acoustic distance:
\begin{equation}
\mathcal{S}_{\mathrm{retr}} = \mathrm{TopK}_{(u_i,y_i)\in\mathcal{D}}\big(-d_{\mathrm{ac}}(u^*,u_i)\big),\quad K\in\mathbb{N}.
\end{equation}

Importantly, transcript-based semantic information is not used for retrieval ranking. Instead, transcripts are directly incorporated into the ICL prompt together with the corresponding audio segments and acoustic descriptors. This design ensures that demonstration selection is grounded in affective acoustic similarity, while semantic reasoning is handled by the LLM during inference.

\subsection{AI-driven Workflows}
\firstpar{The individual components introduced in Sections~3.1 and~3.2 are integrated into a unified AI-driven workflow for multimodal sentiment analysis in immersive collaborative environments. The workflow is designed to provide a coherent transition from data acquisition and representation learning to synthetic annotation and evaluation.}

At the system level, collaborative interactions take place within simulated environments, such as virtual reality setups, which serve as controlled yet ecologically valid experimental platforms. During these interactions, speech signals are recorded in multi-speaker settings, forming the primary data modality considered in this study.

The recorded audio data are subsequently processed to obtain structured representations, including automatic speech transcription, temporal segmentation, and the extraction of acoustic features such as pitch, loudness, intensity, and speaking rate. Pretrained acoustic models and lexicon-based methods are used to derive preliminary affective cues, providing complementary continuous and categorical representations.

These intermediate representations are then incorporated into the LLM-based annotation stage through acoustically informed ICL. By conditioning inference on a small set of enriched few-shot examples selected through acoustic similarity, the system generates synthetic sentiment labels at the segment level. This design enables scalable annotation while reducing reliance on manual labeling.

\section{RESULTS}
\firstpar{This section presents a comparative evaluation of the proposed acoustically informed ICL framework for synthetic sentiment annotation. We first analyze the impact of prompt construction by comparing vanilla few-shot ICL with a retrieval-guided, acoustically informed ICL strategy. We then examine how the proposed ICL approach interacts with different annotation paradigms, including audio-based, lexicon-based, and pretrained text-based sentiment models.}

To ensure a fair and consistent comparison, all methods are evaluated on the same set of 794 segment-level samples, comprising all transcribed speech segments produced by all players within a single VR session. Performance is assessed using normalized confusion matrices and quantitative evaluation metrics, as reported in Tables~\ref{tab:vanilla_vs_gt}--\ref{tab:per_class} and illustrated in Figures~\ref{fig:vanilla_vs_gt}--\ref{fig:xlm_vs_icl}.

\subsection{Few-shot Vanilla ICL vs Acoustically Informed ICL}
\begin{figure}[H]
    \centering
    \includegraphics[width=0.7\textwidth]{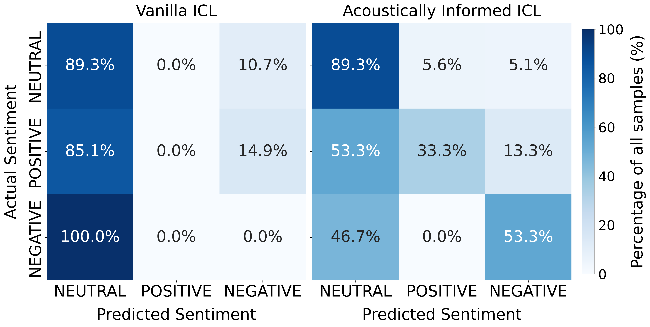}
    \caption{Confusion matrices comparing ground truth sentiment labels with predictions from (left) vanilla few-shot ICL and (right) the retrieval-guided acoustically informed ICL using ground truth-based similarity.}
    \label{fig:vanilla_vs_gt}
\end{figure}

\firstpar{We first examine the effect of prompt construction by comparing vanilla few-shot ICL, which uses randomly sampled demonstrations, with an acoustically informed ICL variant that retrieves a small set of demonstrations using ground-truth-based similarity (Fig.~\ref{fig:vanilla_vs_gt}). The results show that prompt design strongly determines whether the LLM can recognize non-neutral affective states.}

In the vanilla ICL setting (Fig.~\ref{fig:vanilla_vs_gt}, left), predictions collapse almost entirely into the neutral class. The model correctly identifies neutral segments (89.3\%), but fails to recognize the positive class and misclassifies all negative segments as neutral (100.0\% neutral for actual negative). This behavior indicates that randomly selected few-shot prompts do not provide sufficient, discriminative context for the model to separate minority affective states from neutral speech.

\begin{table}[H]
\caption{Performance comparison between vanilla few-shot ICL and acoustically informed ICL.}
\label{tab:vanilla_vs_gt}
\centering
\fontsize{9}{11}\selectfont
\begin{tabular}{|l|c|c|c|c|}
\hline
 & Acc. & Prec. & Recall & F1 \\
\hline
$ICL^{Vanilla}$ & 0.82 & 0.31 & 0.30 & 0.30 \\
\hline
$ICL^{GT}$ & 0.85 & 0.46 & 0.59 & 0.49 \\
\hline
\end{tabular}
\end{table}

In contrast, the retrieval-guided acoustically informed ICL (ground-truth-based retrieval) substantially improves discrimination of non-neutral classes (Fig.~\ref{fig:vanilla_vs_gt}, right). For positive segments, the correct prediction rate increases to 33.3\%, and for negative segments the model achieves 53.3\% correct negative predictions, reducing the systematic ``all-neutral'' failure mode. This improvement is also reflected in the summary metrics: macro-F1 increases from 0.30 to 0.49, while recall rises from 0.30 to 0.59 (Table~\ref{tab:vanilla_vs_gt}). Although overall accuracy changes only slightly (0.82 to 0.85), the macro metrics confirm that the gain comes from improved minority-class recognition rather than from the dominant neutral class.

Overall, this experiment demonstrates that retrieval-guided prompt construction---combined with richer, acoustically grounded context---can fundamentally change the LLM's behavior from neutral-dominant collapse to meaningful separation of positive and negative affect.

\subsection{Retrieval-Guided Acoustically Informed ICL}

\subsubsection{Baseline Annotation Performance}
\firstpar{We next examine three baseline annotation paradigms: an audio-based model (wav2vec 2.0~\cite{ref1}), a lexicon-based text method (NRC-VAD~\cite{ref5}), and a pretrained text-based classifier (XLM-Roberta~\cite{ref11}). Table~\ref{tab:overall_performance} summarizes their overall performance, and Table~\ref{tab:per_class} reports per-class results for the positive and negative classes.}

A key methodological note is that both wav2vec 2.0 and NRC-VAD originally output continuous affective representations (valence, arousal, dominance), i.e., numerical scores in a three-dimensional space. To enable direct comparison with categorical sentiment outputs, these continuous representations are mapped into a simplified three-class scheme (positive, neutral, negative) using predefined mapping rules. While this discretization necessarily removes fine-grained affect information, it ensures that all paradigms are evaluated under the same categorical benchmark.

\begin{figure}[H]
    \centering
    \includegraphics[width=0.7\textwidth]{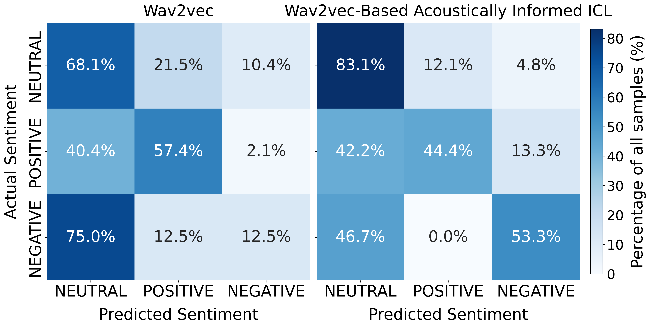}
    \caption{Confusion matrices comparing ground truth sentiment labels with predictions from (left) the wav2vec-based baseline and (right) the retrieval-guided acoustically informed ICL using wav2vec-based similarity.}
    \label{fig:wav2vec_vs_icl}
\end{figure}

Across baselines, overall accuracy is largely driven by the dominance of neutral samples and therefore provides a limited view of model quality. Consistent with the confusion matrices (Fig.~\ref{fig:wav2vec_vs_icl}--\ref{fig:xlm_vs_icl}), all baselines exhibit recurring confusion between neutral and non-neutral classes. Among them, wav2vec and NRC-lexicon achieve comparable macro-F1 scores (0.35 and 0.32 in Table~\ref{tab:overall_performance}, respectively) but show very weak sensitivity to negative affect, with near-zero F1 and recall for the negative class (Table~\ref{tab:per_class}), highlighting their limitations in detecting rare emotional expressions in conversational speech. XLM-Roberta performs slightly better overall (macro-F1 = 0.39) and shows improved balance on minority classes, yet still suffers from substantial neutral--positive confusion, likely due to domain mismatch between written-text training data and spoken team interaction language.

\subsubsection{Effect of Acoustically Informed In-Context Learning}
\firstpar{Applying retrieval-guided acoustically informed ICL consistently improves performance across all three annotation paradigms (Tables~\ref{tab:overall_performance}--\ref{tab:per_class}), with the largest gains occurring in minority-class recognition, particularly for negative sentiment.}

For the wav2vec-based setup, macro-F1 increases from 0.35 to 0.47 (Table~\ref{tab:overall_performance}). The primary improvement is negative-class sensitivity: F1(N) rises from 0.04 to 0.25 and Rec.(N) increases from 0.13 to 0.53 (Table~\ref{tab:per_class}). This trend is also visible in Figure~\ref{fig:wav2vec_vs_icl}, where the ICL-enhanced variant reduces the baseline tendency to absorb negative segments into neutral and improves correct negative predictions.

For the NRC-VAD-based setup, macro-F1 improves from 0.32 to 0.45 (Table~\ref{tab:overall_performance}). Negative-class recognition increases markedly (F1(N): 0.07 to 0.24; Rec.(N): 0.06 to 0.47 in Table~\ref{tab:per_class}). These gains indicate that retrieval-guided ICL can compensate for the absence of prosodic information in lexicon-driven annotation by providing multimodal context through retrieved examples.

\begin{figure}[H]
    \centering
    \includegraphics[width=0.7\textwidth]{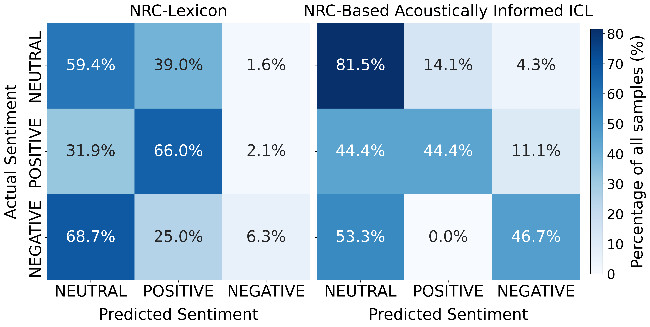}
    \caption{Confusion matrices comparing ground truth sentiment labels with predictions from (left) the NRC-VAD lexicon-based baseline and (right) the retrieval-guided acoustically informed ICL using NRC-VAD-based similarity.}
    \label{fig:nrc_vs_icl}
\end{figure}

\begin{figure}[H]
    \centering
    \includegraphics[width=0.7\textwidth]{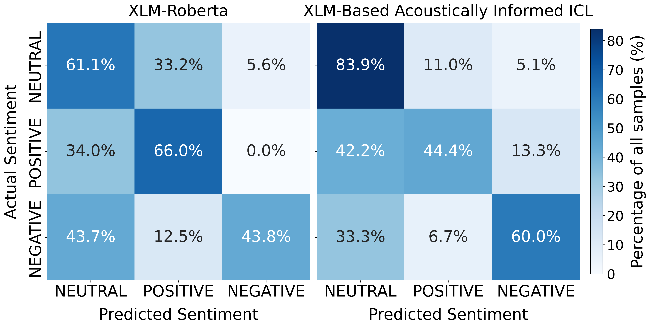}
    \caption{Confusion matrices comparing ground truth sentiment labels with predictions from (left) the XLM-Roberta-based baseline and (right) the retrieval-guided acoustically informed ICL using XLM-Roberta-based similarity.}
    \label{fig:xlm_vs_icl}
\end{figure}

For the XLM-Roberta-based setup, macro-F1 increases from 0.39 to 0.48 (Table~\ref{tab:overall_performance}). Negative-class recall rises from 0.44 to 0.60 and F1(N) improves from 0.22 to 0.27 (Table~\ref{tab:per_class}), consistent with the improved diagonal structure visible in Figures~\ref{fig:nrc_vs_icl}--\ref{fig:xlm_vs_icl}. Given that the text classifier baseline is already relatively strong, the improvements suggest that retrieval-guided ICL acts as a form of contextual adaptation to the characteristics of spoken interaction data.

A consistent trade-off across paradigms is that positive recall decreases (from 0.57/0.66/0.66 to 0.44 across wav2vec/NRC/XLM in Table~\ref{tab:per_class}), while F1(P) increases. This indicates that the ICL-enhanced system becomes more conservative in predicting positive sentiment, reducing false positives at the cost of missing some positive instances. In contrast, improvements for the negative class are both large and consistent, which is particularly important because negative affect is the most challenging and sparsely represented class.

\begin{table}[H]
\caption{Overall performance comparison between baseline methods and acoustically informed ICL.}
\label{tab:overall_performance}
\centering
\fontsize{9}{11}\selectfont
\begin{tabular}{|l|c|c|c|c|}
\hline
 & Acc. & Prec. & Recall & F1 \\
\hline
\emph{wav2vec} & 0.66 & 0.37 & 0.46 & 0.35 \\
\hline
$ICL^{w2v}$ & 0.80 & 0.44 & 0.60 & 0.47 \\
\hline
\emph{NRC} & 0.59 & 0.37 & 0.44 & 0.32 \\
\hline
$ICL^{NRC}$ & 0.79 & 0.43 & 0.58 & 0.45 \\
\hline
\emph{XLM} & 0.61 & 0.40 & 0.57 & 0.39 \\
\hline
$ICL^{XLM}$ & 0.81 & 0.45 & 0.63 & 0.48 \\
\hline
\end{tabular}
\end{table}

\begin{table}[H]
\caption{Per-class performance comparison between baseline methods and acoustically informed ICL.}
\label{tab:per_class}
\centering
\fontsize{9}{11}\selectfont
\begin{tabular}{|l|c|c|c|c|}
\hline
 & F1 (P) & Rec. (P) & F1 (N) & Rec. (N) \\
\hline
\emph{wav2vec} & 0.23 & 0.57 & 0.04 & 0.13 \\
\hline
$ICL^{w2v}$ & 0.26 & 0.44 & 0.25 & 0.53 \\
\hline
\emph{NRC} & 0.17 & 0.66 & 0.07 & 0.06 \\
\hline
$ICL^{NRC}$ & 0.24 & 0.44 & 0.24 & 0.47 \\
\hline
\emph{XLM} & 0.19 & 0.66 & 0.22 & 0.44 \\
\hline
$ICL^{XLM}$ & 0.28 & 0.44 & 0.27 & 0.60 \\
\hline
\end{tabular}
\end{table}

Taken together, the results demonstrate that retrieval-guided acoustically informed ICL is a robust enhancement mechanism across fundamentally different annotation paradigms. The largest relative improvements are observed for wav2vec and NRC-VAD, where the baselines struggle most with minority affective states. For XLM-Roberta, improvements remain consistent but are smaller in magnitude, reflecting its stronger baseline capacity.

Importantly, the gains are best reflected by macro-level metrics and per-class negative performance rather than accuracy. This confirms that the proposed workflow improves sensitivity to non-neutral affective deviations and reduces class-imbalance-driven distortions, making it a practical and scalable approach for synthetic sentiment annotation in immersive team-based VR interaction settings.

\section{CONCLUSIONS}
\firstpar{This work presents a data-efficient and scalable framework for synthetic ground truth generation in speech-based sentiment annotation within immersive virtual reality environments. Experimental results show that a vanilla few-shot ICL approach is insufficient for reliable affective labeling, exhibiting a strong bias toward neutral predictions and limited sensitivity to minority emotional states. By incorporating acoustic information and structured feature representations into a retrieval-guided ICL paradigm, the proposed approach substantially improves annotation robustness and mitigates the effects of class imbalance.}

Beyond individual model comparisons, we introduce an end-to-end AI-driven workflow that automates the sentiment annotation process, spanning representation learning, example retrieval, in-context inference, and evaluation. For controlled and interpretable comparison, this study focuses on a three-class sentiment setting comprising neutral, positive, and negative affective states. While this simplified label space facilitates systematic evaluation, the proposed framework is not inherently restricted to it and can be extended to more fine-grained emotional taxonomies through its retrieval-based and modality-aware design.

Across audio-based, lexicon-based, and pretrained text-based annotation paradigms, the proposed framework consistently outperforms vanilla ICL, demonstrating its general applicability. Although the generated annotations are not intended to replace expert-labeled ground truth, they reliably capture deviations from neutral affective states and are therefore well suited for large-scale analysis scenarios where manual annotation is impractical.

Finally, the proposed methodology is not limited to speech data. Its underlying principles---retrieval-guided ICL, modality-specific grounding, and data-efficient annotation---are transferable to other modalities used in affective computing and psychology, including physiological signals, behavioral measures, and multimodal interaction data. This work thus provides a foundation for unified synthetic ground truth generation to support multi-modal, team-level affective analysis in complex interactive systems.

\section*{ACKNOWLEDGMENTS}
Support by the Technology Transfer Center for Data Science and Autonomous Systems in Landsberg am Lech and the Bavarian Collaborative Research Program (BayVFP), funding line Digitalization, for the project DaTeam-VR is kindly acknowledged.


\begin{thebibliography}{11}
\fontsize{9}{11}\selectfont
\bibitem{ref7} I. A. Castiblanco Jimenez, E. C. Olivetti, E. Vezzetti, S. Moos, A. Celeghin, and F. Marcolin, ``Effective affective EEG-based indicators in emotion-evoking VR environments: An evidence from machine learning,'' \emph{Neural Computing and Applications}, vol. 36, pp. 22245--22263, 2024.

\bibitem{ref6} N. Gao, M. S. Rahaman, W. Shao, and F. D. Salim, ``Investigating the reliability of self-report data in the wild: The quest for ground truth,'' in \emph{Proc. ACM International Joint Conference on Pervasive and Ubiquitous Computing (UbiComp Adjunct)}, pp. 237--242, 2021.

\bibitem{ref2} M. Ihori, T. Yamane, N. Kawata, N. Makishima, T. Tanaka, S. Suzuki, S. Orihashi, and R. Masumura, ``Few-shot personalization via in-context learning for speech emotion recognition based on speech-language model,'' \emph{arXiv preprint arXiv:2509.08344}, 2025.

\bibitem{ref8} M. Mosbach, T. Pimentel, S. Ravfogel, D. Klakow, and Y. Elazar, ``Few-shot fine-tuning vs. in-context learning: A fair comparison and evaluation,'' in \emph{Findings of the Association for Computational Linguistics: ACL 2023}, Toronto, ON, Canada, pp. 12284--12314, 2023.

\bibitem{ref9} A. Radford, J. W. Kim, T. Xu, G. Brockman, C. McLeavey, and I. Sutskever, ``Robust speech recognition via large-scale weak supervision,'' in \emph{Proc. 40th International Conference on Machine Learning (ICML 2023)}, Honolulu, HI, USA, pp. 28492--28518, 2023, [Online]. Available: \url{https://doi.org/10.48550/arXiv.2212.04356}.

\bibitem{ref10} A. H. Liu, A. Ehrenberg, L. A. Lo, C. Denoix, C. Barreau, G. Lample, J.-M. Delignon, K. R. Chandu, P. von Platen, and P. R. Muddireddy, ``Voxtral,'' \emph{arXiv preprint arXiv:2507.13264}, 2025.

\bibitem{ref1} J. Wagner, A. Triantafyllopoulos, H. Wierstorf, M. Schmitt, F. Burkhardt, F. Eyben, and B. W. Schuller, ``Dawn of the transformer era in speech emotion recognition: Closing the valence gap,'' \emph{IEEE Transactions on Pattern Analysis and Machine Intelligence}, vol. 45, no. 9, pp. 10745--10759, 2023.

\bibitem{ref4} S. M. Mohammad, ``Sentiment analysis: Automatically detecting valence, emotions, and other affectual states from text,'' in \emph{Emotion Measurement}, 2nd ed., H. L. Meiselman, Ed. Cambridge, U.K.: Woodhead Publishing, 2021, pp. 323--379.

\bibitem{ref3} S. M. Mohammad and P. D. Turney, ``Crowdsourcing a word--emotion association lexicon,'' \emph{Computational Intelligence}, vol. 29, no. 3, pp. 436--465, 2012.

\bibitem{ref5} S. M. Mohammad, ``NRC VAD lexicon v2: Norms for valence, arousal, and dominance for over 55k English terms,'' 2025, doi: 10.48550/arXiv.2503.23547.

\bibitem{ref11} C. Lalk, K. Targan, T. Steinbrenner, J. Schaffrath, S. Eberhardt, B. Schwartz, A. Vehlen, W. Lutz, and J. Rubel, ``Employing large language models for emotion detection in psychotherapy transcripts,'' \emph{Frontiers in Psychiatry}, vol. 16, Art. no. 1504306, 2025.
\end{thebibliography}
\end{document}